# THEORETICAL CONSIDERATIONS ON THE PROPERTIES OF ACCRETING MILLISECOND PULSARS

Lorne A. Nelson
Physics Department, Bishop's University,
Lennoxville, QC Canada J1M 1Z7
lnelson@ubishops.ca

AND

Saul Rappaport
Department of Physics and Center for Space Research
Massachusetts Institute of Technology, Cambridge, MA 02139
sar@mit.edu

ABSTRACT

We examine a number of evolutionary scenarios for the recently discovered class of accretion-powered millisecond X-ray pulsars in ultracompact binaries, including XTE J0929-314 and XTE J1751-305, with orbital periods of 43.6 and 42.4 minutes, respectively. We focus on a particular scenario that can naturally explain the present-day properties of these systems. This model invokes a donor star that was either very close to the TAMS (i.e., main-sequence turnoff) at the onset of mass transfer or had sufficient time to evolve during the mass-transfer phase. We have run a systematic set of detailed binary evolution calculations with a wide range of initial donor masses and degrees of (nuclear) evolution at the onset of mass transfer. In general, the models whose evolutionary tracks result in the best fits to these ultracompact binaries start mass transfer with orbital periods of $P_{orb} \sim 15$ hr, then decrease to a minimum orbital period of $\lesssim 40$ minutes, and finally evolve back up to about 43 minutes. We present the results of detailed evolutionary calculations for these systems, as well as interior profiles of the donor stars at the current epoch. We find that the initial properties of the donor star (i.e., mass, state of chemical evolution, metallicity), and the exact mode of orbital angular momentum losses during the binary's evolution) do not have to be fine tuned in order to reproduce the observed properties. We also carry out a probability analysis based on the measured mass functions of XTE J0929-314 and XTE J1751-305, and combine this with the results of our binary evolution models to establish estimates of the current properties of these systems. We find that the donor stars currently have masses in the range of $\sim 0.012 - 0.025$ $M_\odot$, and radii of $\sim 0.042 - 0.055$ $R_\odot$, and that these radii are likely to be factors of $\sim 1.1 - 1.3$ times larger than the corresponding radii of zero-temperature stars of the same mass and chemical composition. According to the evolutionary scenario proposed in this paper, the interiors of the donors are largely composed of helium (as opposed to carbon and oxygen), and the surface hydrogen abundances are almost certainly less than 10% (by mass). The orbital period derivative of these systems is very likely to be positive, i.e., $\dot{P}_{orb}/P_{orb} > 0$, with typical values in the range $\sim 3 \times 10^{-10} - 2 \times 10^{-8}$ yr$^{-1}$. Long-term average values of the mass-transfer rate are expected to be $\sim 10^{-11} - 3 \times 10^{-10}$ $M_\odot$ yr$^{-1}$. The evolutionary models, in conjunction with the probability analysis, support the hypothesis that X-ray irradiation is having a minimal effect on enhancing the radius of the low-mass donor and may not even have had a significant effect on the prior evolution of the binary system. We also show how, in the context of this same basic evolutionary scenario, we can model the properties of the SAX 1808.4-3658 binary millisecond X-ray pulsar system ($P_{orb} = 2$ hr). Finally, we point out that if the proposed scenario to explain the ultracompact systems is correct, then these binaries truly link (as evolutionary cousins) systems that (i) evolve to become wide binary millisecond pulsars containing low-mass helium dwarfs (e.g., PSR B1855+09), and (ii) those ordinary accretion-powered, low-mass X-ray binaries in which the hydrogen-rich donors are slowly reduced to planetary masses.

*Subject headings:* — binaries: close — low-mass x-ray binaries — pulsars: millisecond — stars: evolution — stars: mass loss — stars: low mass — stars: hydrogen depleted

## 1. INTRODUCTION

Many of the properties of low-mass, close, interacting binaries (such as Cataclysmic Variables [CVs] and Low-Mass X-ray Binaries [LMXBs]) are reasonably well understood and the formation and evolution of these systems have been investigated in considerable detail (see, e.g., Rappaport, Verbunt, & Joss 1983 [RVJ]; Hameury et al. 1988; Podsiadlowski, Rappaport, & Pfahl 2002 [PRP]). The well established Roche-lobe overflow model (RLOF) describing the evolution of CVs (see, e.g., Warner 1995, and references therein) has also been applied to LMXBs. Both types of systems contain accreting compact objects and their orbital periods tend to be very short (typically between 1 and 10 hours). However several systems with orbital periods of less than one hour have now been discovered. These include the CVs: V485 Cen (59.0 min), GP Com (46.5 min), CP Eri (28.7 min), V803 Cen (26.9 min), HP Lib (18.6 min), CR Boo (24.5 min), AM CVn (17.1 min), and ES Cet (10.3 min) (see, e.g., Podsiadlowski, Han, & Rappaport 2002 and references therein); and the LMXBs: 4U 1915-05 (50.0 min), 4U 1626-67 (41.4 min), 1850-0846 (20.5 min), and 4U 1820-30 (11.4 min) (see, e.g., PRP and references therein). As shown by Nelson, Rappaport, & Joss (1986 [NRJ]; Pylyser & Savonije 1988, 1989; PRP), these ultrashort orbital period systems must contain donor stars that are moderately to severely hydrogen depleted.

Very recently two ultrashort period binaries containing millisecond X-ray pulsars were discovered with the *Rossi X-Ray Timing Explorer* (RXTE) bringing the total number of these accretion-powered millisecond pulsars to three. The X-ray transient XTE J1751-305 has the shortest orbital period with a value of 42.4 minutes and a measured mass function of $1.3 \times 10^{-6}$ $M_\odot$ (Markwardt & Swank 2002a,b), while XTE J0929-314 has an orbital period of 43.6 minutes and the smallest mass function ever measured of $2.7 \times 10^{-7}$ $M_\odot$ (Galloway et al. 2002). The pulse periods of the two pulsars are 2.5 and 5 milliseconds, respectively. Since the mass functions of both of these systems are extremely small, it is likely that both binaries have donors whose masses are only a few hundredths of a solar mass (see, also, Bildsten 2002). XTE J0929-314 may exhibit an H$\alpha$ emission line in its spectrum (Castro-Tirado et al. 2002), although this claim has not yet been independently confirmed. The combination of these properties provides us with a unique opportunity to test our theoretical understanding of the evolution of this type of system, predict the detailed properties of the donor stars, and evaluate the possible effects of X-ray irradiation on the structure of the donors.

There are two basic and competing scenarios for the formation of the ultracompact binary millisecond X-ray pulsars. The first involves a white dwarf (WD) of mass $\sim 0.2 - 0.7$ $M_\odot$ that fills its critical equipotential surface (Roche lobe) and commences mass transfer at very short orbital periods ($\sim 0.7 - 3$ minutes with a neutron star (NS), and then evolves back up to a period of $\sim 43$ min. We label this the "WD–NS" scenario. The second involves a normal donor star of mass $\sim 1 - 3$ $M_\odot$ that begins to transfer mass to the NS after it has just evolved off the main sequence, but before it has become a subgiant. Such systems can evolve to very short orbital periods, e.g., $\gtrsim 5$ min. We denote this as the "TAMS–NS" scenario. We briefly describe these two competing models here, but focus on the TAMS–NS model for the remainder of the paper.

The WD–NS scenario (see, e.g., Rasio, Pfahl, & Rappaport 2000) involves a neutron star in a wide orbit with an intermediate mass secondary (donor) star; for example, its mass could be $M_2 \simeq 2 - 6$ $M_\odot$. Presumably, the evolution to this point involved a prior common envelope phase where the secondary star spiraled into, and ejected, the envelope of the primary (the primary being the progenitor of the NS; see, e.g., Bhattacharya & van den Heuvel 1991, and references therein). If the secondary star evolves to fill its Roche lobe in a wide orbit (i.e., $\sim 10 - 1000$ days), the subsequent mass transfer onto the NS may be dynamically unstable, leading to a second common envelope phase. The result of this second common envelope phase could be either a He WD of mass $\sim 0.2 - 0.45$ $M_\odot$ or a CO WD of mass $\sim 0.55 - 0.7$ $M_\odot$ in a close orbit (i.e., $P_{orb} \lesssim 1$ day) with the NS. If gravitational radiation losses can cause the orbit to shrink substantially on timescales of less than a Hubble time, the WD would fill its Roche lobe at orbital periods of typically between $\sim 0.7 - 3$ min. The subsequent mass transfer from the WD to the NS, driven by gravitational radiation losses, if dynamically stable (see, e.g., Yungelson et al. 2002), would be very rapid, and the orbit would quickly expand to longer periods (see, e.g., Rappaport et al. 1987). After some $\sim 10^8$ yr, $P_{orb}$ could increase to a value of $\sim 43$ min., and the residual mass of the WD would be $\sim 0.01 M_\odot$ (e.g., Rasio et al. 2000). As Bildsten (2002) has shown, a CO WD would have to have non-negligible entropy to explain the properties of either XTE J0929-314 or XTE J1751-305, while even a He WD would have to have a hot interior to match the inferred radius of the companion to XTE J1751-305. The disadvantages of the WD-NS scenario include: (1) the requirement for a NS in a wide orbit about an intermediate mass donor after a common envelope phase; (2) tidal heating to provide significant internal energy to the WDs, and an unknown mechanism for maintaining asynchronism between the WD rotation and orbital motion (see, e.g., Rasio, Pfahl & Rappaport [2000], and references therein); and (3) H would not be expected in the spectrum - if it has indeed been detected (Castro-Tirado et al. 2002). Finally, a distinct advantage of this scenario is that it may naturally account for the overabundance of Ne which has been observed in a number of other ultracompact X-ray binaries (Schulz et al. 2001; Juett, Psaltis, & Chakrabarty 2001).

According to the TAMS–NS scenario, mass exchange occurs between a normal $\sim 1 - 3$ $M_\odot$ donor star and a NS around a time when most, or all, of the H has been depleted in the stellar core (or if the donor star is significantly hydrogen depleted by the time that it has been stripped of most of its mass). The initial orbital period of the binary at the onset of mass transfer would be in the range of $10 - 20$ hours. It has been shown by a number of authors (e.g., Nelson et al. 1986; Pylyser & Savonije 1989; PRP; Nelson, MacCannell & Dubeau 2002 [hereafter NMD]) that: (1) the mass transfer is stable; and, (2) the resultant minimum orbital period can reach values as short as $P_{\min} \sim 5$ min. (see, e.g., Fig. 15 of PRP). This is much shorter than the often discussed (see, e.g., Paczyński & Sienkiewicz 1981; Rappaport, Joss & Webbink 1982 [RJW]; Hameury et al. 1988; Kolb, King & Ritter 1998; Schenker, Kolb, & Ritter 1998; Howell, Nelson & Rappaport 2001 [HNR]) minimum orbital period of $\sim 80$ min. – which holds only for stars with substantial H abundances in their cores (i.e., donors that have not undergone significant nuclear evolution). The advantages of this model include: (1) only one common envelope phase is required; (2) as we show in this work, the initial conditions and input parameters do not have to be finely tuned; (3) the values of $P_{orb}$ when mass transfer commences (e.g., $10 - 20$ hrs) are a natural outcome of the first common envelope phase; and, (4) some H can remain in the donor star once it reaches a $P_{orb} \simeq 43$ minutes. Two disadvantages are that if an overabundance of Ne is eventually discovered in the millisecond pulsar systems with ultrashort orbital periods, then there would be no natural explanation for this anomaly, and, (2) there may be some difficulty in obtaining such a high percentage of ultracompact systems among the observed low mass

X-ray binaries in the Galaxy (i.e., 4 of ∼ 60 systems with a measured $P_{orb}$, [Ritter & Kolb 1998, and updated information]).

As we discuss in this paper (see also Pylyser & Savonije 1988; 1989; PRP) donor stars that commence mass transfer to their companion NS when the orbital period is longer than a certain critical value (i.e., ∼ 10 − 20 hr, depending on the mass of the donor) produce systems that evolve to long periods (of the order of days, and longer). By contrast, if the donor star is not significantly evolved at the start of mass transfer (or does not have the opportunity to become significantly evolved), the systems containing this type of donor will develop into close binaries that approach a minimum orbital period of ∼ 70 − 80 min. (see, e.g., RJW; HNR). In between these two cases it is possible to have donor stars that commence mass transfer when the hydrogen content in their cores has been nearly or completely depleted, but before the star has developed a sizable He core. It is these transition systems which lie near the "bifurcation" limit that can evolve quite naturally to become ultrashort period systems; it is their evolution that will be the subject of this paper.

Specifically, we show how a binary system, consisting of a donor that is sufficiently hydrogen depleted at the start of mass transfer (in particular by the time that it reaches orbital periods of approximately one hour), can evolve to even shorter periods and ultimately become a system with properties similar to those of XTE J0929-314 and XTE J1751-305. Once mass transfer commences, it can easily reach the observed state in ≲ 5 billion years. We emphasize that this result is a natural consequence of the standard Roche lobe overflow (RLOF) model, and that the assumed prescription for magnetic stellar wind (MSW) braking does not have to be fine tuned in order to produce models with the requisite orbital periods. For this to occur, we have found that low-mass donors ($1.0 \lesssim M/M_\odot \lesssim 1.2$) must be very close to TAMS at the onset of mass transfer (metal-poor donors can be slightly less evolved). Higher mass donors (up to 2.5 $M_\odot$) can be much less evolved at the onset of mass transfer but must have sufficient time during the mass-transfer phase to burn significant amounts of hydrogen in their cores in order to attain ultrashort orbital periods. We find that donors with masses initially ≳ 3.5 $M_\odot$ are dynamically unstable against mass transfer. [We caution that the exact mass leading to instability depends strongly on the mass of the neutron star and the assumed mode of systemic (non-conservative) mass loss.] We also show how, in the context of this same basic scenario, we can model the SAX 1808.4-3658 binary millisecond X-ray pulsar system ($P_{orb}$ = 2 hr; Wijnands & van der Klis 1998; Chakrabarty & Morgan 1998).

In §2 of this paper we describe the input physics that was used to carry out the stellar evolution calculations. We also describe the computation of the evolutionary sequences as well as the range of initial conditions that was investigated. Representative evolutionary tracks are also presented. In §3 we discuss the observed properties of XTE J0929-314 and XTE J1751-305 and describe the range and types of initial conditions that allow us to reproduce these properties. In §4 we examine how the observational properties, combined with the binary evolution models, constrain the inferred mass, hydrogen abundance, and radius of the donor star. We also comment on the importance of X-ray irradiation on the present state of the donor. Our conclusions are summarized in the final section.

## 2. EVOLUTIONARY CALCULATIONS

All of the binary evolution calculations were carried out using the Lagrangian-based Henyey method. The basic code has been described in several papers (see, e.g., Nelson, Chau & Rosenblum 1985; NMD) and has been extensively tested. The major modifications are due primarily to improvements in the input physics. In particular, we use the OPAL opacities (Iglesias & Rogers, 1996) in conjunction with the low-temperature opacities of Alexander & Ferguson (1994), and the Hubbard & Lampe (1969) conductive opacities. Great care has been taken to ensure that each of these opacities blends smoothly across their respective boundaries of validity. Our treatment enforces continuity of the respective first-order partial derivatives over the enormous range of the independent variables (i.e., density, temperature, and chemical composition) that are needed to fully describe the properties of the donors.

Since the evolutionary sequences required that the mass of the donor sometimes be reduced to values of less than ten Jovian masses ($\simeq 0.01\ M_\odot$), it was necessary to use an equation of state (EOS) that adequately describes the physics of matter at low temperatures and high densities. We thus employed the Saumon, Chabrier & Van Horn (1995) [SCVH] EOS for hydrogen and helium in conjunction with our own EOS for the heavier elements. The version of their hydrogen EOS that we used incorporated the plasma phase transition associated with pressure ionization. The SCVH EOS takes into account Coulombic interactions and exchange effects. These corrections to a 'perfect gas' EOS can have a very significant effect on the inferred radius of the donor and thus on its theoretical orbital period. Moreover, pressure ionization is an extremely important consideration in determining the detailed structure of the interiors of low-mass stars. We found that the treatment of pressure ionization could have a very significant impact on the computed properties of all low-mass models ($\lesssim 0.3\ M_\odot$). The SCVH EOS does not cover all of the ranges of $T$ and $P$ of interest. In the high-temperature regime ($\log T \geq 7.0$) we used our own equation of state (arbitrarily relativistic and degenerate electrons and weak Coulombic interactions). At extremely low pressures ($\log P \leq 4.0$) we used a previously derived EOS that includes the effects of molecular hydrogen but excludes the very small contributions due to non-ideal effects. The atmospheres were calculated using the prescription of Dorman, Nelson & Chau (1989).

It was assumed that gravitational radiation and MSW braking were responsible for orbital angular momentum losses. We used the RVJ parameterization of the Verbunt-Zwaan (1981) braking law (i.e., $\dot{J} \propto R_2^\gamma$ where $R_2$ is the radius of the donor star and $\gamma$ is a dimensionless number). We have previously investigated a wide range of values for the parameter $\gamma$ and also adjusted the magnitude of the strength of the magnetic braking (see HNR for a detailed discussion of this issue). For the present investigation, we set $\gamma$ equal to values between 3 and 4. The exact value of the radius of gyration of the donor star was determined directly from its com-puted density profile. We also considered the cases of: (i) continuous magnetic braking [MB-C]; and, (ii) interrupted magnetic braking [MB-I] (see RVJ for more details). For the latter case, magnetic braking was stopped when the donor became fully convective. For the MB-C case, we preferred braking laws that were not very strong for orbital periods of approximately one hour. We know that this constraint must be correct for hydrogen-rich CVs for otherwise the theoretical minimum orbital period would be considerably longer than the observed one ($\approx 80$ minutes). We caution, however, that the actual magnitude of magnetic braking is quite uncertain and may be different for LMXBs than for CVs.

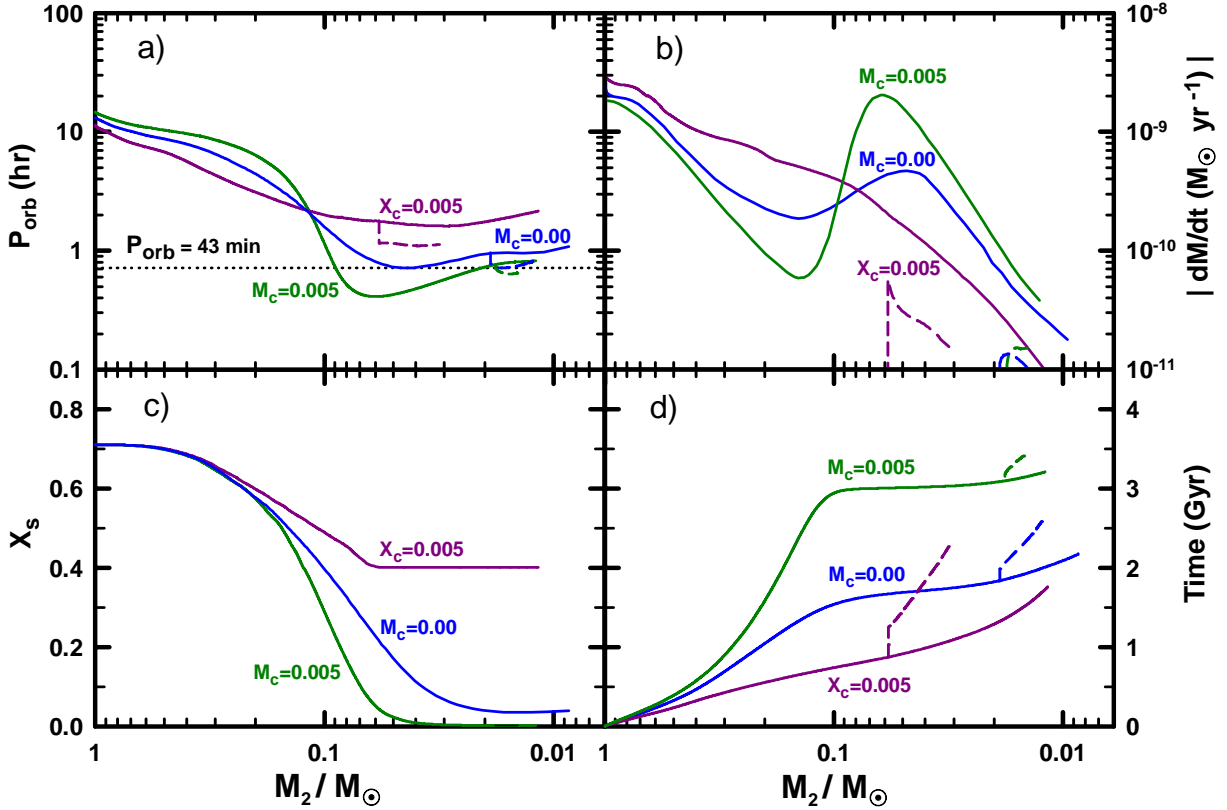

FIG. 1.— Calculated evolution of an initially evolved 1 $M_\odot$ donor in orbit with a neutron star expressed as a function of the donor's mass in solar units ($M_2/M_\odot$). Panel a) shows the calculated orbital period ($P_{orb}$) in hours as a function of mass for three initial chemical profiles of the donor. For all cases the solid curves denote MB-C with $\gamma = 3$ and the dashed curves correspond to MB-I (cessation of magnetic braking after the donor becomes fully convective). The purple curve (labeled $X_c = 0.005$) corresponds to an initial 1 $M_\odot$ donor whose central hydrogen abundance is $X_c = 0.005$ at the onset of mass transfer. The blue curve (labeled $M_c = 0.00$) corresponds to a donor whose central hydrogen abundance has just reached zero at the onset of mass transfer. The green curve (labeled $M_c = 0.005$) corresponds to the case for which the donor has developed a 0.005 $M_\odot$ helium core. The approximate orbital periods of XTE J0929-314 (43.6 minutes) and XTE J1751-305 (42.4 minutes) are denoted by a dotted line corresponding to 43 minutes. Panel b) shows the evolution of the mass transfer rate (in $M_\odot$ yr$^{-1}$) for each respective case. Panel c) illustrates the evolution of the surface hydrogen abundance $X_s$ of the donor for the three cases described above. Panel d) relates the elapsed evolutionary time (since the onset of mass transfer) to the mass of the donor for the three different initial conditions.

We calculated a large number of evolutionary tracks corresponding to different donor masses and degrees of chemical (i.e., nuclear) evolution at the onset of mass transfer. Specifically, we started with donor masses of between $M_2 = 1$ $M_\odot$ and 2.5 $M_\odot$ and initial central hydrogen mass fractions of $0 \leq X_c \leq 0.71$. For the cases corresponding to zero central hydrogen, we considered stars that had helium core masses in the range of $0 \leq (M_c/M_\odot) \leq 0.05$. The initial hydrogen abundance (by mass) on the ZAMS was taken to be equal to 0.71, the mixing length to the pressure scale-height ratio ($l/H_p$) was set equal to 1.5, and the metallicity was taken to be approximately solar ($Z = 0.02$). It is interesting to note that XTE J0929-314 is located at a high Galactic latitude (Galloway et al. 2002). While the binary system may have been born in the galactic plane and subsequently found its way to high latitudes as the result of a supernova kick, it is also possible that it is associated with the Galactic bulge population. If this latter scenario is correct, then a lower than solar metallicity would be appropriate for this system. We have investigated this possibility and find that it has little impact on our conclusions.

The initial mass of the neutron star was taken to be $M_1 = 1.4$ $M_\odot$ and the mass transfer was assumed to be non-conservative. We arbitrarily set the mass-capture fraction to be $|\dot{M}_1|/|\dot{M}_2| = 0.5$ (see PRP for a discussion of this choice) and assumed that the matter lost from the system carried away a specific angular momentum equal to that of the neutron star (i.e., fast Jeans' mode). The actual mass capture fraction does not appreciably alter our conclusions.

Sample results of our evolutionary calculations for $\gamma = 3$ are presented in Figures 1 and 2. The evolution of the orbital period ($P_{orb}$), mass transfer rate ($|\dot{M}_2|$), the donor's surface hydrogen abundance ($X_s$), and elapsed time after the onset of mass transfer ($t$) are shown as a function of the mass of the donor ($M_2$) for initial masses of 1 $M_\odot$ and 1.5 $M_\odot$, respectively. In Figure 1 three cases, corresponding to an initial central hydrogen abundance of $X_c = 0.005$ and helium core masses of $M_c = 0.00 M_\odot$ and $0.005 M_\odot$, are displayed. Other initial conditions corresponding to larger values of $X_c$ do not allow the binary system to evolve to ultrashort orbital periods; donor stars with larger initial helium core masses result in the formation of donors that eventually become sub-giants and the binary evolves to long orbital periods (i.e., those above the bifurcation limit). The donors in these systems eventually detach from their Roche lobes and start to cool, thereby becoming helium degenerate dwarfs [He WDs] (see, e.g., JRL, Pylyser & Savonije 1986; Sarna et al. 2000; PRP; NMD]. An excellent example of a binary that has evolved in this way is the binary millisecond pulsar PSR B1855+09. The mass of the companion to the pulsar has been precisely measured to be $0.258 M_\odot$ to within $\sim 0.02 M_\odot$ (van Kerkwijk et al. 2000) and its effective temperature is consistent with that expected for a cooling He WD.

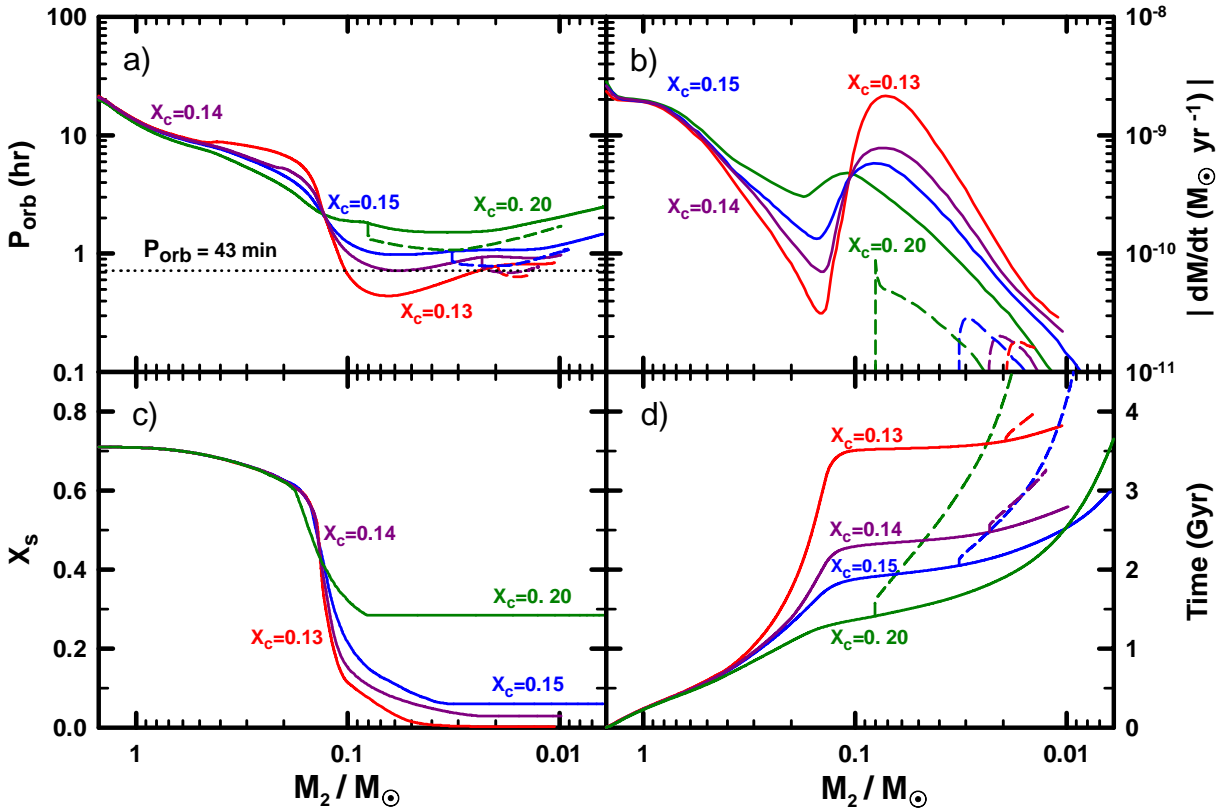

FIG. 2.— Calculated evolution of an initially evolved 1.5 $M_\odot$ donor in orbit with a neutron star expressed as a function of the donor's mass in solar units ($M_2/M_\odot$). All other descriptors are the same as in the caption of Figure 1, except that for this case, the central hydrogen mass fractions (at the onset of mass transfer) are $X_c = 0.13$ (red curve), 0.14 (purple curve), 0.15 (blue curve), and 0.20 (green curve).

In Figure 2 four cases corresponding to an initial donor mass of 1.5 $M_\odot$ and initial central hydrogen abundances of $X_c = 0.20, 0.15, 0.14,$ and $0.13$ are shown. Smaller values of $X_c$ resulted in the formation of long-period binaries wherein the donor ultimately became a He WD. Larger values of $X_c$ did not allow the binary system to evolve to ultrashort orbital periods. For each of the cases illustrated in Figures 1 & 2, the value of the magnetic braking exponent $\gamma$ was set equal to 3 and MB-C was assumed. We also investigated the evolution of donors with initial masses as large as 2.5 $M_\odot$. They too can yield systems with orbital periods as small as 43 minutes as long as the donors have had a chance to evolve and become significantly hydrogen depleted during the mass-transfer phase of the evolution. Donors that are too evolved at the onset of mass transfer, ultimately end up as He WDs or HeCO WDs (i.e., degenerate dwarfs containing helium, or carbon/oxygen) in wide binary orbits. We found that donors whose masses were $\gtrsim 3.5 M_\odot$ were dynamically unstable against mass transfer (see also PRP).

The evolution of the internal properties (i.e., density and temperature) of an initial one-solar-mass donor that has just finished burning all of its hydrogen at its center (i.e., $M_c = 0$) at the onset of mass transfer is shown in Figure 3. A similar plot corresponding to $X_c = M_c = 0.005 M_\odot$ at the start of mass transfer is shown in Figure 4. [Note that these two sets of initial conditions correspond to the second and third cases (respectively) shown in Figure 1.] Each curve shows the variation in density ($\rho$) and temperature ($T$) throughout the interior of the donor for several different masses as it evolves from a 1 $M_\odot$ star down to the lowest masses that were computed ($\sim 0.01 M_\odot$). The mass of the donor corresponding to each curve is labeled in the figures. The dashed lines, labeled $\Gamma = 1$ and $F_{1/2} = 1$, have been plotted in order to illustrate the importance of Coulombic coupling and electron degeneracy (respectively) in the calculation of the interior structure of these models. Specifically, $\Gamma$ is the plasma parameter which is a measure of the ratio of the Coulombic energy of the gas to its thermal energy. Note that $\Gamma = 1$ separates the strongly-coupled regime (to the upper left of the dashed line) from the weakly-coupled regime. The dependence of $\Gamma$ on $\rho$ and $T$ was calculated based on a number-weighted average for a two-component plasma with $X = 0.71$ and $Y = 0.29$. The other line shows the values of $\rho$ and $T$ for which the Fermi integral of index 1/2 is equal to one. This line serves to separate the weakly electron-degenerate regime (region to be lower right of the line) from the intermediate (and strongly) degenerate regime.

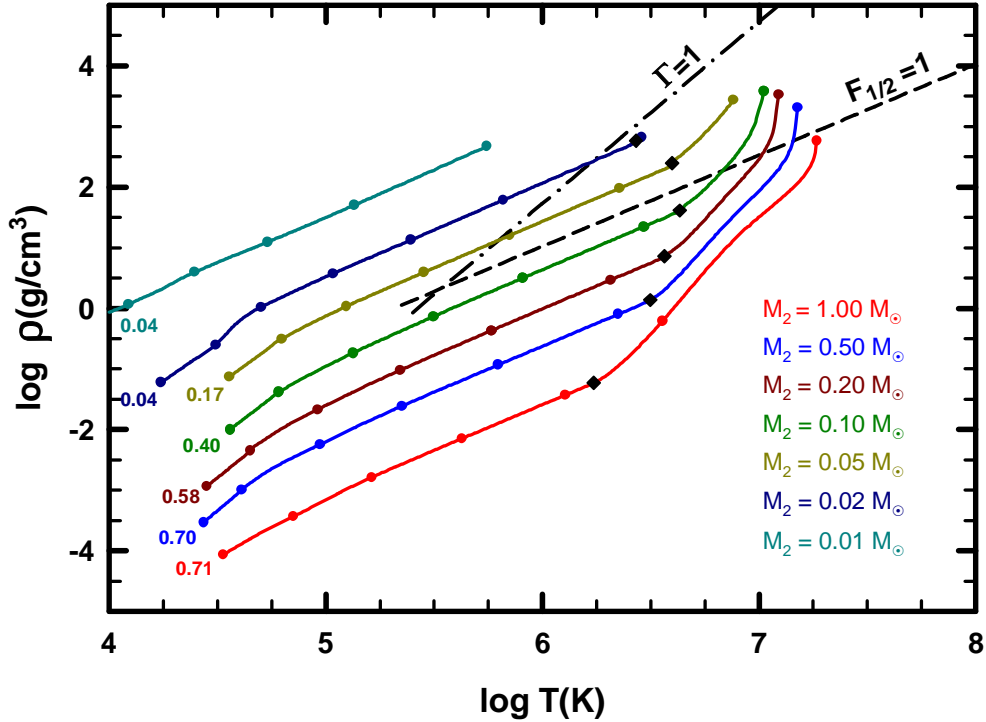

FIG. 3.— The run of density ($\log \rho$) versus temperature ($\log T$) of an initial 1 $M_\odot$ donor with $X_c = 0$ that has undergone mass transfer under the assumption that the system is experiencing a MSW with MB-C with $\gamma = 3$. Each curve corresponds to a different mass (the mass of each model is shown in the figure). The dots on each curve denote equal logarithmic intervals in the mass fraction (i.e., $\log(1 - m_r/m)$). The right most dot at the end of each curve corresponds to an interior mass fraction of 0% (i.e., the stellar center), the next corresponds to 90% of the mass being interior to that point, and so forth, with the left most dot corresponding to 99.999% of the stellar mass being interior to that point. Next to this dot is the value of $X_s$ (i.e., the hydrogen abundance at the surface). The diamond symbol on each curve denotes the location in the stellar interior where the transition from a radiative/conductive core to a convective envelope occurs. The dashed lines, labeled $\Gamma = 1$ and $F_{1/2} = 1$ for a composition of $X = 0.71$ and $Y = 0.29$, have been plotted in order to illustrate the importance of Coulombic coupling and electron degeneracy (respectively).

The dots on each curve denote equal logarithmic intervals in the mass fraction (i.e., $\log[1 - m_r/m]$). The right-most dot at the end of each curve corresponds to an interior mass fraction of 0% (i.e., the stellar center), the next corresponds to 90% of the mass being interior to that point, and so forth, with the left-most dot corresponding to 99.999% of the stellar mass being interior to that point. The diamond symbol on each curve denotes the location in the stellar interior where the transition from a radiative/conductive core to a convective envelope occurs. Note that the convective envelope moves deeper into the interior of the star as its mass is decreased. As a consequence of this effect, the envelope (and hence surface) becomes increasingly hydrogen-poor as the hydrogen-depleted gas from the deep interior is mixed into the envelope. The actual values of $X_s$ for each mass are labeled beside the left-most points of each curve in Figures 3 and 4. We clearly see a significant decrease in $X_s$ as the convective envelope moves inwards. For both cases shown in Figures 3 and 4, the donor star becomes fully convective at a mass of slightly less than 0.02 $M_\odot$ (this is in contrast to a hydrogen-rich ZAMS star which becomes fully convective at a mass of $\sim 0.32$ $M_\odot$). Also note in Figure 4 that there is a significant 'jump' in the spacing of the $\log \rho - \log T$ curves between the 0.10 $M_\odot$ and 0.05 $M_\odot$ models (contrast this difference with that seen for the same models of Figure 3). This is due to the varying abundances of hydrogen in the envelope and atmosphere of the donor stars. Specifically, for the $M_c = 0.0$ case, the value of $X_s$ is 0.40 for the 0.10 $M_\odot$ donor and 0.17 for the 0.05 $M_\odot$ donor. By contrast, those same values are 0.29 and 0.02 (respectively) for the $M_c = 0.005$ $M_\odot$ case. The large (relative) decrease in the hydrogen abundance of the 0.05 $M_\odot$ model necessarily implies a significantly different internal structure.

The plots of the interior properties also indicate that the cores of donors whose masses are in the range of $0.2 \lesssim M_2/M_\odot \lesssim 1.0$ are somewhat isothermal. In addition, we see that when the mass of the donor is $\lesssim 0.05$ $M_\odot$, Coulombic coupling is an important contributor to the EOS (area to the left of $\Gamma = 1$ in Figures 3 and 4) and hence to the overall determination of the properties of the donor. Moreover, since $\Gamma$ is composition dependent, the corresponding $\Gamma = 1$ line for helium rich interiors (appropriate for all of our low-mass donors) should be shifted considerably to the right of (but parallel to) the ones shown in Figures 3 and 4. For example, under the assumption that $X = 0$, the $\Gamma = 1$ line should intersect the top edge of the figures (corresponding to $\log \rho = 5$) at $\log T = 7.425$. Finally we note that even the lowest mass donors still have considerable thermal energy (as is evidenced by their lack of isothermality) and are far from solidifying (crystallization occurs for $\Gamma \simeq 170$).

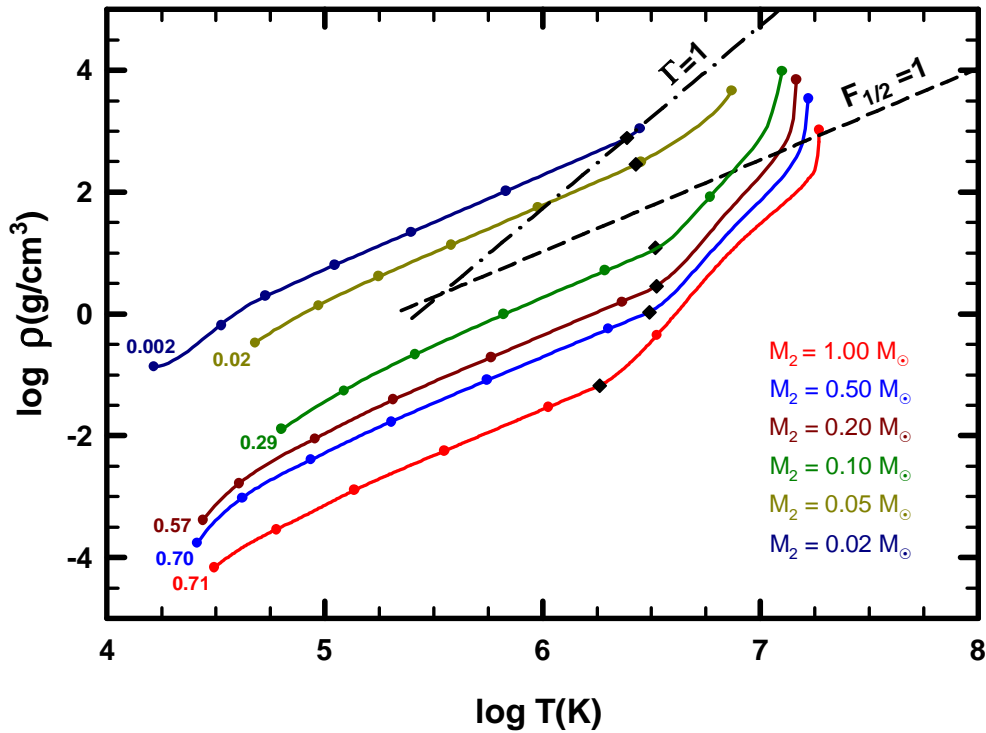

Fig. 4.— The run of density ($\log \rho$) versus temperature ($\log T$) of an initially one-solar-mass donor with $M_c = 0.005$ that has undergone mass transfer under the assumption that the system is experiencing MB-C with $\gamma = 3$. All other descriptors are the same as in the caption of Figure 3.

### 3. THEORETICAL PROPERTIES OF THE PULSAR SYSTEMS

Our evolutionary calculations confirm the claims of NRJ [see also Pylyser & Savonije 1988, 1989; PRP] that only extremely hydrogen-depleted donors can produce ultrashort period systems. According to NRJ, a chemically homogeneous (i.e., convective) donor must have a hydrogen abundance of no more than a few percent in order to evolve to values of $P_{orb} \lesssim 1$ hour. Our evolutionary sequences show that for almost all of the initial conditions considered, the donors in XTE J0929-314 and XTE J1751-305 should have a current value of $X_s \lesssim 0.15$ (values closer to, or consistent with, zero are favored). More importantly, we find that there is a substantial volume of parameter space that will yield systems with properties similar to those of the two millisecond pulsar binaries. For example, conservative or non-conservative evolutions with a wide range of braking parameters (e.g., amplitudes and braking exponents) will produce ultrashort period systems. Initial donor masses in the range of $1 \lesssim (M_{2,i}/M_\odot) \lesssim 2.5$ are possible and no value of the metallicity is excluded. However, once a particular choice is made for the donor's initial mass, its metallicity, and the mode of angular momentum dissipation, tight constraints are necessarily imposed on the initial state of chemical (i.e., nuclear) evolution of the donor. These constraints must ensure that the donors become extremely hydrogen depleted by the time that their mass has been reduced to a few hundredths of a solar mass.

As a representative example of this type of evolution consider the case of a 1 $M_\odot$ donor with an initial helium core mass of $M_c = 0.01 \ M_\odot$. We find that a value of $P_{orb} \simeq 43$ minutes is reached twice during the subsequent evolution when $X_s \simeq 0.23$ and $M_2 \simeq 0.097 \ M_\odot$, and when $X_s \simeq 0.0002$ and $M_2 \simeq 0.02 \ M_\odot$ ($\gamma = 3$ and MB-C assumed). In the former case the binary had not yet reached the minimum value of $P_{orb}$, while in the latter case the system has evolved past that point. Although these initial conditions do yield plausible models they do not correspond to the best fit of the observed properties of either binary. In the former case, the value of $M_2$ is most probably far too large based on the measured mass functions (see §4 below) while in the latter case the value of $X_s$ may be too small[1] for the case of XTE J0929-314 if the detection of hydrogen in the spectrum (Castro-Tirado et al. 2002) is confirmed. The initial conditions (for a one-solar-mass donor) that seem to best fit the data correspond to an initial value of $M_c = 0$ (i.e., when hydrogen is first depleted in the center of the donor). For the $\gamma = 3$ (MB-I) case, we find that $P_{orb} = 43$ minutes when $X_s \simeq 0.04$ and $M_2 \simeq 0.017\,M_\odot$ while for $\gamma = 4$ and with both MB-I and MB-C the values are $X_s \simeq 0.03$ and $M_2 \simeq 0.014\,M_\odot$. The secular mass-transfer rates for all three of these systems (at $P_{orb} = 43$ min) are between $\sim 0.8 - 2.0 \times 10^{-11}\,M_\odot \text{yr}^{-1}$. It is difficult to use the calculated secular (long-term average) mass-transfer rates as a diagnostic tool to test the validity of the models since both systems are transient sources, each seen on only one occasion. It is possible to have higher rates but the values of $M_2$ would have to be larger. The properties of a representative sample of systems that attain $P_{orb} = 43$ minutes are shown in Table 1. Given that the smaller values of $M_2$ ($\sim 0.02\,M_\odot$) are more probable based on the measured mass function (see §4), we predict that the value of $X_s$ is likely to be very small ($\lesssim 0.05$).

---
[1] It should be noted that Williams & Ferguson (1982) claim that in the accretion disks of CVs, the Balmer emission lines are comparable in strength to the He I emission lines even when the number ratio of helium to hydrogen is as large as 100. Thus the detection of hydrogen may still be consistent with an extremely small value of $X_s$ (but probably not as small as $X_s = 0.0002$).

To obtain a binary system whose properties correspond to the observed ones, it is clear that a ($1\,M_\odot$, Population I) donor must be a TAMS star at the onset of mass transfer. Similar results can be obtained starting with larger mass donors (e.g., $1.5\,M_\odot$) but in those cases the donors must be slightly less evolved at the onset of mass transfer ($0.12 \lesssim X_c \lesssim 0.15$; see Table 2 for more details). The higher mass stars have shorter nuclear timescales and can thus burn some of their central hydrogen before the binary evolution can cause sufficient mass to be stripped away and thereby effectively extinguish nuclear evolution. Even higher mass stars are good candidates but must be considerably less evolved at the onset of mass transfer. Analogously, a donor star with a smaller metallicity (such as might be observed for Galactic bulge stars) evolves faster than a solar metallicity star and thus would also need to be slightly less evolved at the onset of mass transfer in order to produce a system whose properties match those that are observed (see NMD for further details). For all of these cases, if the donor is to have a very low mass ($\lesssim 0.02\,M_\odot$) then the mass transfer rate is also likely to be low ($\lesssim 10^{-10}\,M_\odot \text{yr}^{-1}$). Finally, we point out that in some of these models it is possible for the minimum period to be as short as 8 minutes! However, it is worth noting that such systems are much less likely to be detected simply because they evolve through this extreme ultrashort period range very quickly.

It is interesting to note that in order to obtain an orbital period of $\sim 43$ minutes and to have any appreciable amount of hydrogen remaining at the donor's surface, the binary should be (for most initial conditions) reasonably close to its minimum orbital period (see, e.g., NRJ). As has been pointed out by a number of authors, (see, e.g., Paczyński & Sienkiewicz 1981, RJW, and Chau & Nelson 1982), the Kelvin-Helmholtz (thermal) timescale must necessarily be approximately equal to the mass-loss timescale near the minimum orbital period (i.e., $\tau_{KH} \simeq \tau_{\dot M}$). Thus the thermal timescale can be estimated from the mass-loss timescale using the data presented in Figures 1b and 2b and the definition of $\tau_{\dot M}$ ($\equiv M_2/|\dot M_2|$). The data indicate that this timescale is usually $\lesssim 2 \times 10^9$ years, thereby implying that the donor still has a significant amount of thermal energy to shed and that it should be substantially larger than its corresponding zero-temperature radius ($R_0$). In fact, our most favored fits for XTE J0929-314 indicate that the donor is likely to be at least $\sim 25\%$ larger than the corresponding $R_0$.

## 4. OBSERVATIONAL CONSTRAINTS

In the previous sections we presented detailed binary evolution models that can reproduce systems with properties similar to those of XTE J0929-314 and XTE J1751-305. As the donor stars in these systems lose mass and evolve, their interior structures are substantially more complicated than simple polytropes (see, e.g., Fig. 3). However, by the time they attain masses of $\lesssim 0.02\,M_\odot$ their chemical composition is approximately uniform and they have only a small residual amount of hydrogen. Furthermore, their structures can be approximated by cold degenerate dwarfs, but with some thermal "bloating". Therefore, in order to gain some further insight into the evolutionary status of the companion stars in these millisecond pulsar systems, we develop an analytic model which makes use of the stellar properties already deduced from the more detailed models discussed in §2 and §3.

We start with the assumption that the donor stars in these systems are currently filling their Roche lobes. We take the volume-averaged radius of the Roche lobe, $R_L$, to be

$$R_L = \frac{2}{3^{4/3}} \left( \frac{M_2}{M_2 + M_1} \right)^{1/3} a \ , \quad (1)$$

where $a$ is the orbital separation and $M_2$ and $M_1$ are the donor and neutron star masses, respectively (Paczyński 1971). If we combine equation (1) for the Roche lobe, which we take to equal the radius of the donor, $R_2$, with the expression for Kepler's Third Law, we find the following well-known relation between the orbital period, $P_{orb}$, the donor mass, and the donor radius:

$$P_{orb} = \frac{9\pi}{\sqrt{2G}} R_2^{3/2} M_2^{-1/2} \ . \quad (2)$$

If there exists a well-defined mass-radius relation for the donor star, then the orbital period can be related directly to the mass of the donor star. For this we utilize a fitting formula due to Eggleton (2003) for the dependence of radius on mass and chemical composition for zero-temperature objects:

$$R_2 \simeq 0.0128(1+X)^{5/3} f \left( \frac{M_2}{M_\odot} \right)^{-1/3} g(M_2; X)\, R_\odot \ , \quad (3)$$

with

$$g(M_2; X) = \left[ 1 - \left( \frac{M_2}{M_{ch}} \right)^{\frac{4}{3}} \right]^{\frac{1}{2}} \times$$
$$\left[ 1 + 3.5 \left( \frac{M_2}{M_p} \right)^{-\frac{2}{3}} + \left( \frac{M_2}{M_p} \right)^{-1} \right]^{-\frac{2}{3}} , \quad (4)$$

$$M_{ch} = 5.76\, \langle Z_N/A \rangle^2 \, M_\odot \simeq 1.44(1+X)^2\, M_\odot \ ,$$

$$M_p = 0.0016\, \langle Z_N/A \rangle^{3/2} \, \langle Z_N^2/A \rangle^{3/4}\, M_\odot$$
$$\simeq 0.00057(1+X)^{3/2} M_\odot \ .$$

Note that the angular brackets denote a mass-weighted average, and $Z_N$ and $A$ are the atomic number and atomic weight, respectively, of each of the chemical constituents of the donor star. The rightmost expressions for $M_{ch}$ and $M_p$ are for objects composed of H and He only. Also $f$ is the thermal "bloating factor" defined as the ratio of the actual donor radius to its zero-temperature radius (i.e., $f = R_2/R_0$). The Eggleton expression does a very good job of matching the radii of the zero-temperature (hydrostatic) models of Zapolsky & Salpeter (1969).

TABLE 1. SUMMARY OF MODEL PROPERTIES ($M_{2,0} = 1.0\ M_\odot$; $P_{orb} = 43$ MIN)

| Initial Conditions[a] | $\left(\frac{M_2}{M_\odot}\right)$[b] | $t$ (Gyr)[c] | $X_s$ | $\langle X\rangle$[d] | $\log(|\dot M_2|)$[e] | $\log\left(\frac{L_2}{L_\odot}\right)$ | $P_{min}$[f] | $\left(\frac{R_2}{R_\odot}\right)$ | $\left(\frac{\dot P_{orb}}{P_{orb}}\right)$[g] | $f$[h] |
|---|---|---|---|---|---|---|---|---|---|---|
| MB-C3; $M_{c,0} = 0.00$ | 0.051 | 1.69 | 0.176 | 0.052 | −9.3 | −3.15 | 41.3 | 0.071 | $-2.8 \times 10^{-8}$ | 2.13 |
| | 0.038 | 1.72 | 0.099 | 0.033 | −9.4 | −3.54 | 41.3 | 0.064 | $+2.2 \times 10^{-8}$ | 1.84 |
| MB-I3; $M_{c,0} = 0.00$ | 0.017 | 2.15 | 0.036 | 0.036 | −10.9 | −4.46 | 43.0 | 0.048 | $\approx 0$ | 1.15 |
| MB-C4; $M_{c,0} = 0.00$ | 0.068 | 3.91 | 0.232 | 0.062 | −9.7 | −2.91 | 34.5 | 0.075 | $-3.3 \times 10^{-9}$ | 2.43 |
| | 0.015 | 4.64 | 0.027 | 0.027 | −10.8 | −4.68 | 34.5 | 0.046 | $+3.0 \times 10^{-10}$ | 1.08 |
| MB-I4; $M_{c,0} = 0.00$ | 0.014 | 4.74 | 0.027 | 0.027 | −11.0 | −4.96 | 40.7 | 0.045 | $-4.6 \times 10^{-10}$ | 1.06 |
| MB-C3; $M_{c,0} = 0.005$ | 0.089 | 2.98 | 0.226 | 0.044 | −9.3 | −2.59 | 24.7 | 0.082 | $-2.0 \times 10^{-8}$ | 2.94 |
| | 0.021 | 3.07 | 0.0019 | 0.001 | −9.9 | −3.83 | 24.7 | 0.051 | $+3.7 \times 10^{-9}$ | 1.34 |
| MB-I3; $M_{c,0} = 0.005$ | 0.018 | 3.13 | 0.0016 | 0.002 | −10.6 | −3.96 | 38.4 | 0.049 | $-2.3 \times 10^{-10}$ | 1.23 |
| MB-C3; $M_{c,0} = 0.01$ | 0.097 | 3.52 | 0.232 | 0.040 | −9.6 | −2.44 | 20.9 | 0.084 | $-2.2 \times 10^{-8}$ | 3.09 |
| | 0.019 | 3.62 | 0.0002 | 0.0002 | −10.0 | −3.89 | 20.9 | 0.050 | $+3.3 \times 10^{-9}$ | 1.28 |
| MB-C3; $M_{c,0} = 0.02$ | 0.151 | 9.60 | 0.422 | 0.023 | −9.9 | −1.18 | 8.0 | 0.097 | $-2.7 \times 10^{-8}$ | 4.22 |
| | 0.014 | 9.71 | 0.000 | 0.000 | −10.2 | −4.31 | 8.0 | 0.045 | $+1.8 \times 10^{-9}$ | 1.07 |

[a] MB-C and MB-I refer to the cases of continuous magnetic braking and interrupted magnetic braking, respectively, and the number immediately following (3 or 4) corresponds to the value of the dimensionless braking exponent $\gamma$. $M_{c,0}$ refers to the mass of the pure helium core (in units of $M_\odot$) that has been formed at the onset of mass transfer.
[b] Note that in some cases two values of $M_2$ are given because there are two times during the evolution when $P_{orb} = 43$ minutes is attained. For the case of MB-I, the value quoted corresponds only to that phase of the evolution after magnetic braking has been halted.
[c] The age is measured from the onset of mass transfer.
[d] Hydrogen abundance averaged over the mass of the donor.
[e] The mass-transfer rate expressed in units of $M_\odot$ yr$^{-1}$.
[f] The value of $P_{min}$ is expressed in minutes and corresponds to the actual minimum orbital period of the evolutionary track.
[g] Normalized orbital period time derivative in units of yr$^{-1}$.
[h] Approximate thermal bloating factor for a homogeneous chemical composition.

The models of Zapolsky & Salpeter were calculated based on an equation of state (EOS; Salpeter & Zapolsky 1967) of zero-temperature matter that included the contributions from arbitrarily relativistic and degenerate electrons as well as many non-ideal effects due to the long-range and short-range interactions among electrons and ions. We have constructed our own zero-temperature models based on the Salpeter & Zapolsky EOS and conclude that the Eggleton expression (given by equations [3] and [4]) is typically accurate to within $\sim 1\%$ over a mass range of $0.003 - 0.1\ M_\odot$.

If we now combine equations (2) and (3) we find

$$P_{\rm orb} = 0.769(1 + X)^{5/2} f^{3/2} \left(\frac{M_2}{M_\odot}\right)^{-1} g(M_2; X)^{3/2},\quad (5)$$

where $P_{\rm orb}$ is in units of minutes. If we knew the value of $X$ and that of the bloating factor, $f$, we could deduce $M_2$ from the measured orbital period alone. As an example, take $P_{\rm orb} = 43$ min, $X = 0$, and $f = 1$; that is, a completely degenerate, hydrogen-exhausted donor. Solving equation (5) for $M_2$ yields $\sim 0.012\ M_\odot$. For relatively small values of $X$, the donor mass scales approximately as $(1 + X)^{5/2} f^{3/2}$ (see eq. (5)).

The measured mass functions $[F(M)]$ of the systems of interest (see §1 for the numerical values) provide another crucial piece of information, but introduce two other unknown quantities into the problem:

$$F(M) = \frac{M_2^3 \sin^3 i}{(M_2 + M_1)^2},\quad (6)$$

namely, the mass of the neutron star ($M_1$) and the orbital inclination angle, $i$. Based on studies of neutron stars in other binary systems (e.g., Joss & Rappaport 1976; Thorsett & Chakrabarty 1999; Barziv et al. 2001) we have a reasonable expectation that $M_1$ is likely to lie in the range of $\sim 1.3 - 1.8\ M_\odot$.

For our first analysis of the donor masses in the two millisecond X-ray pulsar systems, as well as of the quantity $(1 + X)^{5/3} f$, we adopt a mass for the neutron star of $M_1 = 1.4\ M_\odot$. In a subsequent analysis, we consider a wider range of neutron star masses–up to $2\ M_\odot$. For the inclination angle we assume an *a priori* probability for finding the system with an inclination angle $i$ of

$$\frac{dp(i)}{di} = \sin i \ . \qquad (7)$$

Table 2. Summary of Model Properties ($M_{2,0} = 1.5\ M_\odot$; $P_{orb} = 43$ min)

| Initial Conditions[a] | $\left(\frac{M_2}{M_\odot}\right)$[b] | $t$ (Gyr)[c] | $X_s$ | $\langle X \rangle$[d] | $\log(|\dot{M}_2|)$[e] | $\log\left(\frac{L_2}{L_\odot}\right)$ | $P_{min}$[f] | $\left(\frac{R_2}{R_\odot}\right)$ | $\left(\frac{\dot{P}_{orb}}{P_{orb}}\right)$[g] | $f$[h] |
|---|---|---|---|---|---|---|---|---|---|---|
| MB-C3; $X_{c,0} = 0.13$ | 0.102 | 3.50 | 0.120 | 0.031 | -9.2 | -2.27 | 26.4 | 0.089 | $-2.1 \times 10^{-8}$ | 3.25 |
|  | 0.024 | 3.60 | 0.0026 | 0.002 | -9.7 | -3.70 | 26.4 | 0.054 | $+2.8 \times 10^{-9}$ | 1.45 |
| MB-I3; $X_{c,0} = 0.13$ | 0.019 | 3.67 | 0.0020 | 0.002 | -11.1 | -3.89 | 38.4 | 0.050 | $-1.2 \times 10^{-9}$ | 1.29 |
| MB-C3; $X_{c,0} = 0.14$ | 0.059 | 2.38 | 0.075 | 0.029 | -9.2 | -2.95 | 42.9 | 0.072 | $\approx 0$ | 2.35 |
| MB-I3; $X_{c,0} = 0.14$ | 0.021 | 2.68 | 0.029 | 0.029 | -10.7 | -4.03 | 41.5 | 0.052 | $-5.1 \times 10^{-10}$ | 1.30 |
|  | 0.015 | 3.06 | 0.029 | 0.029 | -10.9 | -4.78 | 41.5 | 0.046 | $+3.6 \times 10^{-10}$ | 1.07 |

[a] MB-C and MB-I refer to the cases of continuous magnetic braking and interrupted magnetic braking, respectively, and the number immediately following (3) corresponds to the value of the dimensionless braking exponent $\gamma$. $X_{c,0}$ refers to the value of $X$ at the center of the donor at the onset of mass transfer.
[b] Note that in some cases two values of $M_2$ are given because there are two times during the evolution when $P_{orb} = 43$ minutes is attained. For the case of MB-I, the value quoted corresponds only to that phase of the evolution after magnetic braking has been halted.
[c] The age is measured from the onset of mass transfer.
[d] Hydrogen abundance averaged over the mass of the donor.
[e] The mass-transfer rate expressed in units of $M_\odot$ yr$^{-1}$.
[f] The value of $P_{min}$ is expressed in minutes and corresponds to the actual minimum orbital period of the evolutionary track.
[g] Normalized orbital period time derivative in units of yr$^{-1}$.
[h] Approximate thermal bloating factor for a homogeneous chemical composition.

To a good approximation, for small values of $M_2$

$$M_2 \simeq \frac{F(M)^{1/3}\ M_1^{2/3}}{\sin i} \ . \qquad (8)$$

Note that we do solve equation (6) *exactly* for $M_2$ for the results presented in Figs. 5 and 6. Equations (6) and (7) allow us to determine the cumulative probability that the donor's mass in a particular system has a value $\geq M_2$ (for an assumed value of $M_1$)[2]. A similar type of analysis can be applied to the bloating factor $f$ (although not completely independent of $X$). With the constraint imposed by eq. (5) and knowing $P_{orb}$ for each system, we can also determine the approximate cumulative probability that the quantity $(1+X)^{5/3}f$ is greater than a particular value. To accomplish this, we in essence set $X \approx 0$ in the $g(M_2;X)$ term of eq. (4) by requiring that $M_p = 0.00057\ M_\odot$ (i.e., $Z_N = 2$ and $A = 4$). This is a reasonable approximation given that the $(1+X)^{5/3}$ term dominates over the $g(M_2;X)$ term in the limit of $X << 1$. For example, even if $X$ were as large as 0.1 (which we believe to be unlikely based on the computed models and given the probability constraints on the donor's mass), the relative error in the value of the cumulative probability is at most only a few percent.

[2] Note that for the range of values of $M_1$ considered and based on the measured mass function for XTE J1751-305, all possible values of $i$ are allowed except those for which eclipses would have been observed (no eclipses are observed). For XTE J0929-314 restrictions must be placed on the inclination angle so as to ensure that the H/He donor does not overfill its Roche lobe (otherwise we would require a value of $f$ [$< 1$] which is physically inadmissible). As an example, assuming that $M_1 = 1.4\ M_\odot$, we require $i$ to be $\leq 45°$.

Cumulative probability distributions for $M_2$ for both XTE J0929-314 and XTE J1751-305 are shown in Figure 5a, while the corresponding distributions for the quantity $(1+X)^{5/3}f$ are shown in Figure 5b. In both cases, the adopted mass for the neutron star is 1.4 $M_\odot$. For XTE J0929-314 the lowest allowed donor mass is 0.012 $M_\odot$, while for XTE J1751-305 it is 0.014 $M_\odot$. Figure 5a shows that the 95% confidence upper limits on the masses of the donors of these two systems are $\lesssim 0.048\ M_\odot$ and $\lesssim 0.046\ M_\odot$, respectively. Similarly the 95% confidence limits for $(1+X)^{5/3}f$ are 2.2 for XTE J0929-314 and 2.1 for XTE J1751-305. The lower limit on $(1+X)^{5/3}f$ for the latter system is 1.07 (see Table 3 for a more complete set of data and the summary of an analysis that was carried out for a 2.0 $M_\odot$ neutron star).

In addition to these distributions, we have carried out a *joint* analysis for the compound probability distribution of $M_2$ and $(1+X)^{5/3}f$ under the *assumption* that both systems are essentially "twins". The compound probabilities have been calculated under the assumption that both neutron stars and both donors have the same mass. The *joint* cumulative distributions are superposed on both panels of Figure 5. The 95% confidence limits for a canonical 1.4 $M_\odot$ neutron star on $M_2$ and $(1+X)^{5/3}f$ are:

$$M_2 \lesssim 0.023 M_\odot \qquad (9)$$

$$1 \lesssim (1+X)^{5/3}f \lesssim 1.42 \ . \qquad (10)$$

For $X = 0$, this corresponds to an upper limit on the bloating factor $f$ of $\sim 1.4$, while if we adopt a value of $X = 0.1$ (a reasonable maximum for the hydrogen abundance based on our computed models [see §2]), the upper limit on the bloating factor is ∼ 1.2. A more detailed comparison of these results (for both 1.4 $M_\odot$ and 2.0 $M_\odot$ neutron stars) with the evolutionary models can be found in Table 3.

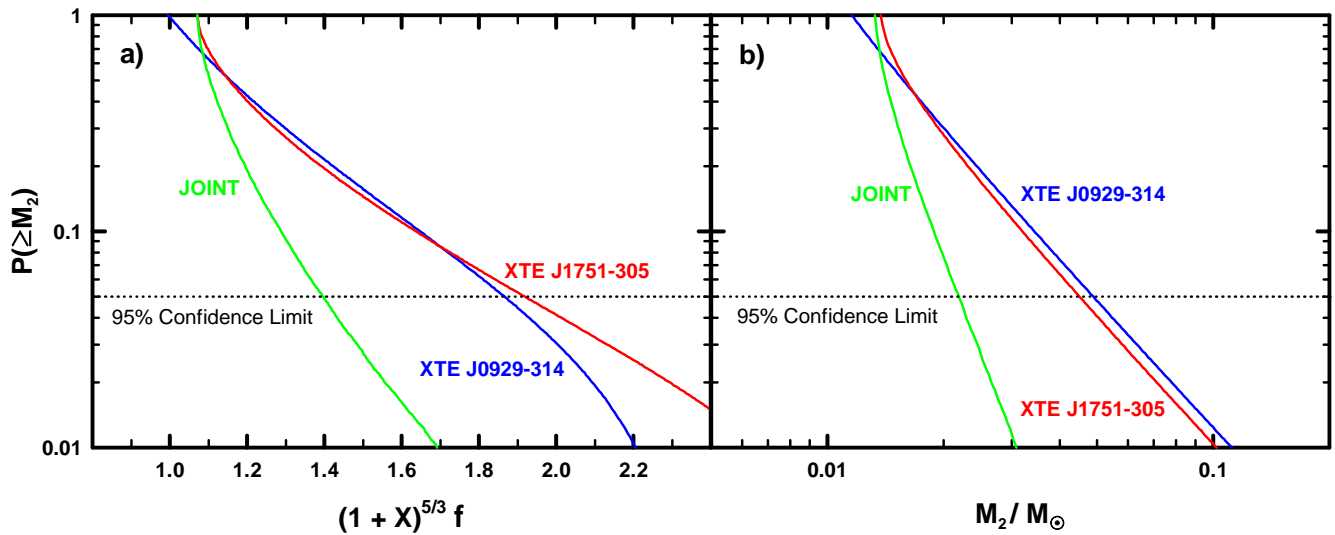

Fig. 5.— Panel a) shows the cumulative probability of finding the donors in XTE J0929-314 (blue curve) and XTE J1751-305 (red curve) with masses larger than a particular value (i.e., $M_2/M_\odot$). The green curve is the "joint probability" that the donors would have at least that mass (under the assumption that the two systems are twins; see the text for details). Panel b) shows the cumulative probability of finding the donors in XTE J0929-314 (blue curve) and XTE J1751-305 (red curve) with particular values of $(1+X)^{5/3} f$, where $X$ is the hydrogen abundance by mass and $f$ is the thermal bloating factor. The green curve corresponds to the "joint probability".

TABLE 3. INFERRED RANGES OF PHYSICAL PROPERTIES

| Property[a] | $M_1$ ($M_\odot$)[b] | J0929-314[c] | J1751-305[c] | Joint[c] | Models[d] |
|---|---|---|---|---|---|
| $M_2$ ($M_\odot$) | 1.4 | 0.012 − 0.048 | 0.014 − 0.046 | 0.012 − 0.023 | 0.013 − 0.06 |
|  | 2.0 | 0.015 − 0.061 | 0.017 − 0.058 | 0.015 − 0.029 | 0.013 − 0.06 |
| $R_2$ ($R_\odot$) | 1.4 | 0.042 − 0.070 | 0.044 − 0.066 | 0.042 − 0.052 | 0.045 − 0.07 |
|  | 2.0 | 0.047 − 0.074 | 0.048 − 0.072 | 0.047 − 0.058 | 0.045 − 0.07 |
| $(1+X)^{5/3} f$ | 1.4 | 1.00 − 2.21 | 1.07 − 2.10 | 1.00 − 1.42 | 1.05 − 2.4 |
|  | 2.0 | 1.12 − 2.54 | 1.21 − 2.42 | 1.12 − 1.62 | 1.05 − 2.4 |

[a] Physical properties include the mass ($M_2$) and radius ($R_2$) of the donor (secondary), and a term $[(1+X)^{5/3} f]$ dependent on a combination of its hydrogen abundance ($X$) and thermal bloating factor ($f$). See the text for more details.
[b] Mass of the neutron-star primary ($M_1$) in units of solar masses.
[c] Approximate 95% confidence intervals for the systems J0929-314 and J1751-305. The interval corresponding to 'Joint' is based on an assumed compound probability for both systems.
[d] Approximate range of properties inferred from our evolutionary models at $P_{orb} = 43$ minutes. Note that only post-minimum-period models (i.e., those with increasing values of $P_{orb}$) are included in this range.

Constraints among the parameters $M_2$, $f$, and $X$ are further explored in Figure 6. In this figure, the relationships among these parameters are solved implicitly using our zero-temperature models that were computed with the 26-parameter EOS of Salpeter & Zapolsky (1967). As mentioned previously, these models yield radii very similar to those predicted by the radius-mass relation given by eqs. (3) & (4). The relationships between $X$ and $M_2$ for constant values of $f$ for XTE J0929-314 and XTE J1751-305 are shown in Figures 6a and 6c, respectively. Note that the region to the upper left of the $f = 1$ curve in the $X − M_2$ plane of both figures is unphysical (assuming that the donors are composed solely of hydrogen and helium). Figures 6b and 6d show the corresponding relations between $X$ and $f$.

The values of $X$, $M_2$, and $f$ can be constrained using statistical inference in conjunction with the measured mass function and a specific value for $M_1$. Assuming that the inclination angle is randomly distributed (as given by equation [7] and given the caveats concerning constraints on $i$), a probability analysis allows us to simultaneously place constraints on the hydrogen abundance and thermal bloating factor of the donor. For illustrative purposes, we take a representative range for the mass of the neutron star to be $1.4 \leq M_1/M_\odot \leq 2.0$. The results of this analysis are shown for XTE J0929-314 and XTE J1751-305 in Figures 6b and 6d, respectively. The solid and dashed curves, defined by the variable $\mathcal{P}$, are probability contours and correspond to the lower and upper limits of our representative range of neutron star masses, respectively. For a given contour, we can state with a degree of confidence equal to $\mathcal{P}$ that the values of $f$ and $X$ must lie to the lower left of the corresponding curves in the $f - X$ plane. For example, at $\mathcal{P} = 90\%$, and taking the mass of the neutron star to be equal to the canonical value of 1.4 $M_\odot$, we can conclude that $f$ for XTE J1751-305 must be less than 1.7. For non-zero values of the hydrogen abundance, the upper limit on $f$ would be even smaller. Even if the donor were fully electron degenerate, the maximum hydrogen abundance that would be permitted is only $\sim 0.4$. If the mass of the neutron star were significantly larger than 1.4 $M_\odot$, then the constraints on $f$ and $X$ would be somewhat less severe. These results are consistent with the ones derived from Figure 5.

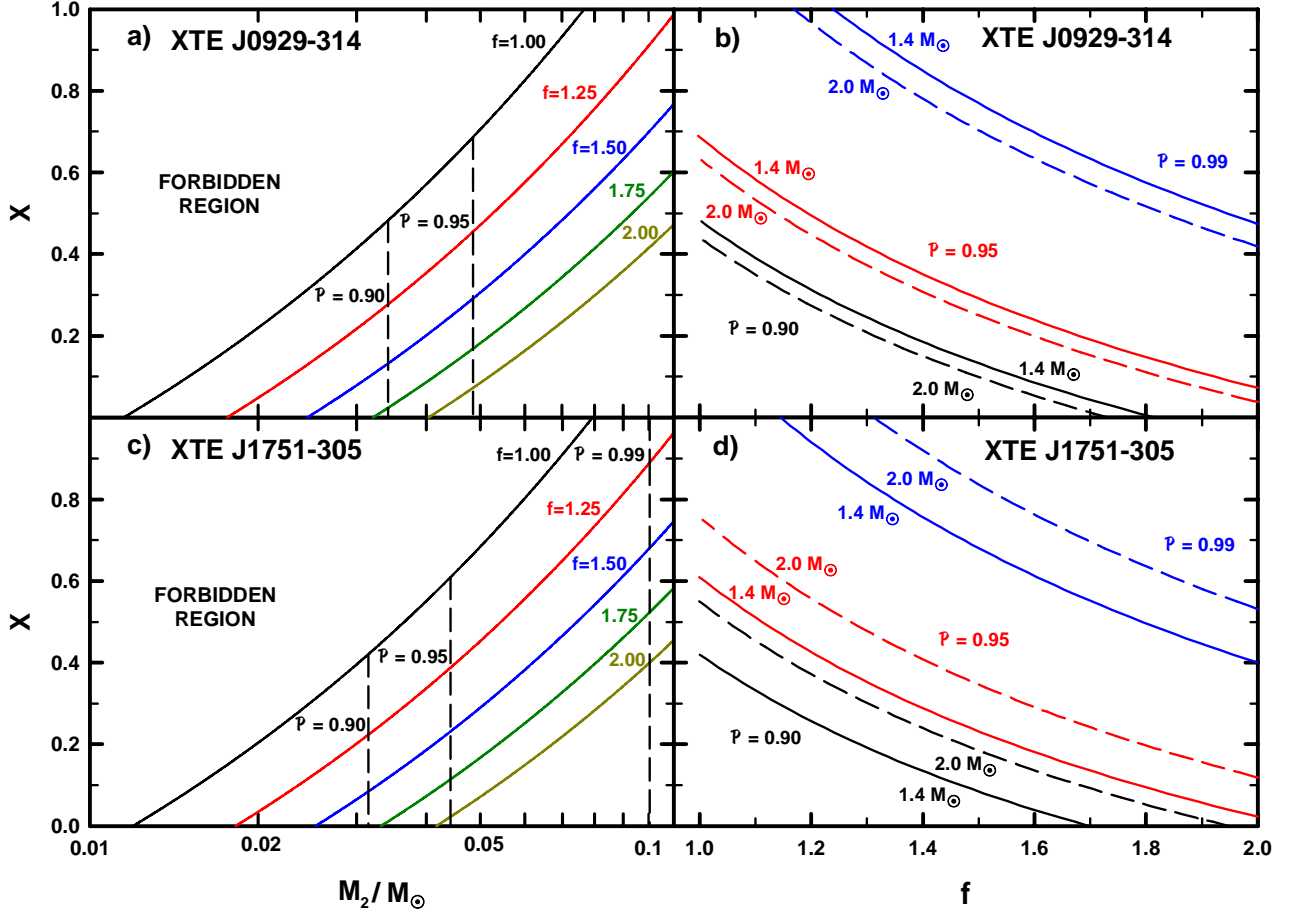

FIG. 6.— Panel a) shows the relationship between the hydrogen mass fraction $X$ and the mass of the donor $M_2$ for fixed values of $f$. Note that the upper left-hand portion of the $X - M_2$ plane is forbidden since values of $f < 1$ are physically inadmissible. Probability contours for the upper limit on the mass of the donor of XTE J0929-314 (based on its measured mass function and assuming a canonical neutron star mass of 1.4 $M_\odot$) are indicated by dashed vertical lines. They denote the degree of confidence $\mathcal{P}$ that can be attached to a particular upper limit (e.g., $\mathcal{P} = 95\%$). Panel b) shows the constraints that can be simultaneously placed on the hydrogen mass fraction $X$ and the thermal bloating factor $f$ for various degrees of confidence ($\mathcal{P}$). The solid curves correspond to an assumed mass of the neutron star equal to 1.4 $M_\odot$ and the dashed curves correspond to a mass of 2.0 $M_\odot$. Allowed values of $X$ and $f$ lie to the lower left of the respective constraint curves. Panels c) and d) are analogous to those of a) and b), respectively, except that the curves have been calculated for XTE J1751-305.

The respective probability contours are also plotted (as dashed lines) on Figures 6a, c for a neutron star whose mass is 1.4 $M_\odot$. Note the strong constraints placed on the donor's mass, its thermal bloating factor, and its hydrogen abundance at a 95% degree of confidence (see also Figure 5). For this degree of confidence (and assuming a 1.4 $M_\odot$ companion), we see that the mass of the donor is in the range of $0.012 \lesssim M_2/M_\odot \lesssim 0.044$. This range of masses is in agreement with several of the evolutionary models discussed in the previous section (see Table 3). To accommodate this mass range we require that the donor be very hydrogen-depleted (typically $\langle X \rangle \lesssim 0.1$ by the time that the orbital period has been reduced to about one hour). The actual values of $f$ for our model donors that are not yet fully mixed (i.e., homogeneous) are tedious to quantify because we would have to calculate individual zero-temperature models with exactly the same internal chemical profiles; but, it is clear that values of $f$ usually do not fall below $\sim 1.1$ (at $P_{orb} = 43$ minutes) unless the donor masses are very low ($\lesssim 0.015 \, M_\odot$; see Tables 1 & 2). Values of $f$ that are significantly larger than unity indicate that the donors have not had enough time to shed their thermal energy (also see the discussion in §2). The cooling timescales of isolated low-mass ($\lesssim 0.1 \, M_\odot$) He WDs are typically more than several billion years (see, Rappaport et al. 1987, Hansen & Phinney 1998). The lowest value of $f$ that we were able to infer from the models (at $P_{orb} = 43$ minutes) was $f \simeq 1.04$ and that corresponded to an evolution driven by gravitational radiation alone (after the cessation of magnetic braking) and the mass of the donor was near or below (depending on the mass of the neutron star) the value allowed by the measured mass functions of the two systems. On the other hand, some of our models had bloating factors that exceeded 3.0; but in those cases, the large mass of the donor made the models highly improbable.

These results are sufficiently constraining that they allow us to comment on the importance of X-ray irradiation on the (present-day) structure of the donors in the two systems. Based on the theoretically predicted mass-transfer rates of $\sim 10^{-10}~M_\odot~{\rm yr}^{-1}$ (see Figures 1b and 2b for donor masses of $\sim 0.02~M_\odot$), we would expect that a copious flux of X-rays should be produced by the accreting neutron star. Assuming an isotropic distribution of radiation and knowing the solid angle subtended by the donor, we can calculate the maximum integrated X-ray luminosity that could be intercepted by the donor. According to our models this intercepted luminosity could be $\gtrsim 1 L_\odot$. This is generally several orders of magnitude higher luminosity than the donor would be radiating if it were not exposed to the X-ray radiation (see Tables 1 & 2). Such a large amount of X-ray irradiation should greatly increase the thermal energy of the outer layers and could possibly lead to a significant enhancement of the bloating factor $f$ (for a detailed analysis of effects of X-ray heating on low-mass companions see, for example, Podsiadlowski 1991; Tavani & London 1993; Hameury et al. 1993; Harpaz & Rappaport 1995; Ritter, Zhang & Kolb 2000, and references therein). As discussed above, the observational evidence indicates that it is not very likely the bloating factor exceeds values of $\sim 1.4$ (see the 'joint analysis' of Figure 5). Moreover, if the detection of hydrogen in XTE J0929-314 is correct, the maximum likely bloating factor would be considerably less. On the other hand, according to our most favored evolutionary models (i.e., those for which $M_2 \lesssim 0.023~M_\odot$ and $|\dot{M_2}| \gtrsim 10^{-10}~M_\odot {\rm yr}^{-1}$), $f$ has typically not decreased to a value much smaller than $\sim 1.3$ – *without* any X-ray heating effects being included. Since the observational upper limit and the theoretical lower limit on the bloating factor are not very different, this implies that X-ray heating is relatively unimportant with respect to enhancing the radius of the donor (at the present epoch). We cannot envision a situation within the context of the RLOF model wherein X-ray irradiation would actually decrease the size of the donor below that inferred from our evolutionary models since X-ray irradiation should lead to the deposition of energy in the atmosphere and envelope of the donor thereby increasing its temperature and inhibiting its ability to cool the core region. Furthermore, there is no need to invoke X-ray heating (except perhaps if the donor is a HeCO WD) to adequately explain all of the observed properties of these two millisecond pulsar binaries.

Assuming that the donor has evolved according to the RLOF model and that X-ray irradiation has not contributed significantly to the bloating of the donor, we must reconcile this result with the potentially large flux of X-rays (e.g., $> 10^{13}$ ergs cm$^{-2}$ sec$^{-1}$) that could be intercepted by the donor. There are two primary reasons why we might expect the *excess bloating* due to X-ray irradiation to be small:

(i) X-ray eclipses are very rare in LMXBs and thus it is plausible that the accretion disks in many LMXBs effectively prevent the X-ray radiation from reaching the donor (see Milgrom [1976] for an early discussion of eclipse probabilities in LMXBs). Since the angle subtended by the donor (with respect to the neutron star) is approximately equal to $0.46(M_2/M_T)^{1/3}$ radians (for $M_2 << M_1$), extremely low-mass donors such as the ones in XTE J0929-314 and XTE J1751-305 would subtend angles of only 5 to 10 degrees. This may be less than the opening angle of the disk. Moreover, the intense X-ray heating experienced by the disk probably produces an extended disk atmosphere (e.g., an "accretion disk corona"; White & Holt 1982) that effectively blocks most of the radiation from directly reaching the donor's surface.

(ii) Because the donor is locked into a synchronous orbit due to the efficiency of tidal interactions, approximately 1/2 of the surface area of the donor will always face the neutron star companion (i.e., the hot-side/cold-side model). If the circulation timescale in the envelope of the donor is sufficiently long (perhaps a large superadiabatic zone exists that effectively prevents the transport of heat inwards and azimuthally), then it might be possible that the effects of the X-ray deposition are minimal (see, e.g., Hameury et al. 1993, Harpaz & Rappaport 1995; and, Ritter, Zhang, & Kolb 2000). Specifically, in that case, the internal heat of the donor could "leak" out of the cold side of the star.

If case (i) is correct, then the results of the evolutionary calculations presented in §2 should be quite accurate. If case (ii) applies, then a more detailed analysis is required. The greatest effect would probably be on the transition mass for which complete convection occurs and this in turn would affect the surface hydrogen abundance. Nonetheless, based on our probability analysis of the empirical data, we expect that X-ray irradiation has a minimal effect on enhancement of the radius of the present-day donors of the two millisecond pulsars.

## 5. SUMMARY AND CONCLUSIONS

There are a number of viable evolutionary scenarios that can lead to the formation of ultracompact binary millisecond pulsars (see, e.g., §1). In this paper we have focused on the TAMS–NS scenario wherein the initial conditions (at the onset of mass transfer) place them close to the bifurcation limit. More specifically, the donor star must convert a sufficient quantity of hydrogen into helium in order to achieve ultrashort periods while maintaining (continuous) Roche-lobe contact throughout their evolutionary histories. If their nuclear evolution exceeds a certain threshold (i.e., puts them above the bifurcation limit), they will eventually evolve to become (non-accreting) wide binary millisecond radio pulsars. Without significant nuclear evolution, however, the binary will evolve with decreasing orbital periods until a period-minimum (of $\sim 80$ min.) is reached and the system will then evolve back to periods approaching two hours. The evolution of these LMXBs then resembles the prototypical evolution of cataclysmic variables (most CVs evolve in this way but seven have orbital periods of less than one hour). Systems starting mass transfer from near the bifurcation limit (i.e., with $P_{\rm orb} \simeq 15$ hr would evolve to become the ultrashort-period binaries described in this paper. The relative probabilities of forming these three different types of systems depends on several factors, and these are discussed in §5.2 below.

Within the context of the TAMS–NS scenario, we have computed detailed binary evolution models for XTE J0929-314 and XTE J1751-305 (see Figs. 1 – 2, and Tables 1 & 2) and have shown that, in the absence of X-ray irradiation, a hydrogen-rich donor can evolve to the observed ultrashort orbital periods as long as the donor star is sufficiently evolved at the onset of mass transfer (or by the time it attains a low mass, i.e., $\lesssim 0.1\ M_\odot$). These models are a very natural extension of the standard main sequence-neutron star pairing that have been invoked to reproduce the observed properties of LMXBs. We have used these models to make specific predictions as to the properties of the donor star at the current epoch (see Figs. 3 – 4) and, in particular, to suggest that the donors will be composed of a small, but possibly observable, fraction of hydrogen. We have also utilized a probability analysis in conjunction with both the measured mass functions and our evolutionary models to show that the donor stars are likely to have masses of $\sim 0.02\ M_\odot$, radii of $\sim 0.05\ R_\odot$, surface hydrogen abundances of $X_s \lesssim 0.1$, and thermal bloating factors of $1.05 \lesssim f \lesssim 1.4$ (see §2 and §3, and Tables 1 and 2). Our evolutionary calculations allow for a wide range of initial masses and metallicities for the donor, and do not require that the exact physical description of the magnetic braking torques (and/or the cessation of MSW braking) on the orbit be fine tuned. The presence of hydrogen would be a key discriminant for determining the validity of the TAMS–NS evolutionary scenario, and a precise measurement constraining its value would be extremely helpful in accelerating theoretical progress.

According to the TAMS–NS scenario it should be noted that while the donors in XTE J0929-314 and in XTE J1751-305 are low-mass objects, they cannot truly be considered as hydrogen (or helium) brown dwarfs, or even as (cold) degenerate dwarfs. Their present-day properties have largely been dictated by: (i) their prior nuclear evolution; and, (ii) the rate at which they were stripped of mass by their neutron star companions. According to our most favored models for the donors in these systems, their current intrinsic luminosities should be $\sim 10^{-4} - 10^{-5}\ L_\odot$ and their radii should be approximately 5% to 40% larger than their respective zero-temperature radii. Detailed profiles of the interiors of the donor stars in selected models at various stages in their evolution are shown in Figs. 3 and 4. The donors should continue to lose mass (and cool) on a very long timescale and, in the process, start to resemble planets. If and when the X-ray phase ends, a millisecond radio pulsar may turn on and start to evaporate the donor (see, e.g., van den Heuvel & van Paradijs 1988). The implications of this possibility need to be examined more carefully.

Based on the properties of our evolutionary models, it is likely that the effects of X-ray heating are small (in terms of significantly enhancing the radii of the two donor stars). There may be several reasons for this; for example, the very small angle subtended by the low-mass donor may be comparable with, or smaller than, the opening angle of the disk. Nevertheless, we plan to investigate the effects of X-ray deposition on the evolution of these extremely low-mass stars in a future paper

(Nelson & Rappaport 2003a). In particular, we would like to ascertain the extent to which efficient convection would be suppressed by X-ray heating and to examine its overall impact on the evolution of the mass-losing donor.

### 5.1. *SAX J1808.4-3658*

A very important system that may be intimately related to the two ultrashort period systems is the accreting millisecond pulsar, SAX J1808.4-3658 ($P_{orb} = 2$ hr). The measured mass function is $3.85 \times 10^{-5}\ M_\odot$ (Chakrabarty & Morgan 1998) and this implies a likely donor mass of $\sim 0.06\ M_\odot$ (see eq. 8). The mass of the neutron star becomes untenable ($\lesssim 1\ M_\odot$) if the donor has a mass smaller than $\sim 0.035\ M_\odot/(\sin i)$. It is interesting to note that some of the properties of the SAX pulsar can be explained using our binary evolutionary sequences. For example, consider the $X_c = 0.005$ case shown in Figure 1 ($M_{2,0} = 1\ M_\odot$). It is clear that when the orbital period equals 2 hours, the mass of the donor is reduced to $\sim 0.09\ M_\odot$ and the corresponding mass-transfer rate is $\sim 3 \times 10^{-10}\ M_\odot\ \mathrm{yr}^{-1}$.

Chakrabarty and Morgan (1998) estimate the mass transfer rate (based on recurrent outbursts) to be $\sim 10^{-11}\ M_\odot\ \mathrm{yr}^{-1}$. This estimate, based on the assumption that all of the mass lost by the donor is accreted by the pulsar (i.e., efficient mass transfer), is a factor of 30 times smaller than the theoretical one. However, as mentioned previously, the value of the long-term (secular) average rate of mass transfer obtained from the models may not be reflective of the rates inferred for transient sources. Figure 2 shows that an even wider range of initial conditions (starting with a $1.5\ M_\odot$ donor) will lead to the formation of a two-hour pulsar system. In this case the current-epoch donor mass is tightly constrained to be about $0.12\ M_\odot$. The mass-transfer rate from the models span an order of magnitude (averaging $\sim 2 \times 10^{-10}\ M_\odot\ \mathrm{yr}^{-1}$). In the worst case scenario, assuming a donor mass of $\sim 0.12\ M_\odot$ and a range of neutron star masses between 1.4 and $2.0\ M_\odot$, we conclude that the inclination angle for SAX J1808.4-3658 would be in the range of $20° \lesssim i \lesssim 30°$. But by adjusting the initial conditions and magnetic braking law, we can reduce the donor mass to a much more probable value of $\sim 0.07\ M_\odot$. A similar claim was made by Ergma & Antipova (1999) who found that a $1\ M_\odot$ star near turn-off could be brought into Roche-lobe contact and evolve to produce a system similar to the SAX pulsar. They also found that the accreted matter should be helium rich (similarly, see our Figures 1c and 2c), and they use this result to explain the short total outburst duration observed for the source.

By contrast, King (2001) argues that the faint transient X-ray sources in the Galactic Center region (of which SAX J1808.4-3658 is one) are likely to be binaries that have evolved as conventional (hydrogen-rich) LMXBs. These systems evolve from orbital periods on the order of hours and reach their minimum orbital periods ($\sim 80$ minutes) before continuing to evolve up to increasing orbital periods (e.g., 2 hours). King claims that the population of observed Galactic Center transients are likely to be post-minimum 'LMXBs' ($P_{orb} \sim 80$ min. to 2 hr) and that this model is consistent with the low mass-transfer rates inferred for these sources. Within the framework of this scenario, he also notes that the elapsed evolution time for SAX J1808.4-3658 would have to be extremely long (i.e., close to the age of the Galaxy). But if this scenario were actually applicable to the SAX pulsar, then the mass of the donor would have to be so small ($\lesssim 0.03\ M_\odot$) that it could almost certainly be ruled out based on the measured mass function (otherwise the mass of the neutron star would have to be exceedingly small).

### 5.2. Expected Population of Ultracompact X-Ray Binaries

An important issue that we have not yet rigorously evaluated concerns the *probability* of creating systems with ultrashort periods within the framework of the TAMS–NS model. This would entail carrying out a detailed binary population synthesis (BPS) study, which is beyond the scope of the present work. In such a study, probabilities are specified for the masses of the primordial stellar components, the initial binary separation, the birthrate frequency, a heuristic description of the physics associated with common envelopes, an assumed distribution of natal kicks for neutron stars, and a model for the angular momentum losses associated with magnetic braking (see, e.g., HNR and Pfahl, Rappaport, & Podsiadlowski 2003 for more details). An accurately performed analysis would provide the relative numbers of systems evolving to become wide binary radio millisecond pulsars, ordinary LMXBs, and ultracompact X-ray binaries, respectively. The contributions to the latter category coming from the TAMS–NS and WD–NS channels would also be derived from the BPS. The current-epoch population of ultracompact binaries involves the relative branching ratios among the various channels described above, as well as the evolution times spent in the ultracompact state relative to the time spent evolving to that state. Finally, such a BPS simulation would also yield the absolute numbers for the spatial density of the ultracompact binaries. Of course reconciling the spatial densities with the observed values would be subject to substantial selection effects.

A good first estimate of the evolutionary 'dwell time' spent in the ultracompact state versus the time spent by the binary evolving to that state can be simply expressed in terms of the mass-loss timescale ($|M_2/\dot{M}_2|$). This quantity is dependent on the response of the donor to mass loss and is typically approximately equal to the timescale associated with the driver of the mass transfer. For $P_{\rm orb} \lesssim 2$ hr angular momentum losses due to magnetic braking are of the same order as those due to gravitational radiation (GR) (or less effective), in which case the mass-loss timescale will be approximately the same as the gravitational radiation timescale:

$$\tau_{GR} \equiv \left|\frac{J_{orb}}{\dot{J}_{GR}}\right| \simeq 3\times 10^7 \left(\frac{P_{orb}}{43\ \rm min}\right)^{\frac{8}{3}} \left(\frac{M_2}{M_\odot}\right)^{-1} \times \left(\frac{M_1}{M_\odot}\right)^{-1} \left(\frac{M_T}{M_\odot}\right)^{\frac{1}{3}} \rm yr. \qquad (11)$$

Equation (11) demonstrates the strong dependence of the timescale on the orbital period, and a somewhat weaker dependence on the mass of the donor (the dependence on $M_1$ and $M_T$ can affect the timescale by at most a factor of two). We infer from eq. (11) that the evolutionary timescale for the two ultracompact millisecond X-ray pulsars, based on a donor mass of $\sim 0.02\ M_\odot$ (and assuming no magnetic braking), is approximately one Gyr. This 'dwell time' is sufficiently large that it should not be an overriding factor in diminishing the probability of detection of systems such as XTE J0929-314 and XTE J1751-305. If magnetic braking dissipated orbital angular momentum by a factor of, for example, 5 times greater than gravitational radiation, then the dwell time would concomitantly be reduced by a factor of 5. We also note from eq. (11) that the corresponding dwell time for even shorter orbital period systems (e.g., $P_{orb} \sim 10$ minutes). would be very short lived (e.g., $\sim 3\times 10^7$ yr). Based on a preliminary inspection of our binary evolution results, we note that higher mass donors ($\gtrsim 1\ M_\odot$) in longer period binaries are probably more favored in the TAMS–NS evolutionary scenario simply because it takes approximately $10^{10}$ years before a $1\ M_\odot$ star can reach the TAMS. Since the total time required for a binary to reach a particular evolutionary state is the sum of this pre-mass-transfer evolution time plus the time elapsed during the subsequent mass-transfer (the latter times being shown in Figs. 1d & 2d), systems born with lower mass donors may not be able to reach the desired state within the age of the Galaxy. Because there are many dimensions of parameter space to consider, the only unbiased way to evaluate the relative probabilities of finding ultracompact binaries among the general population of LMXBs is via a binary population synthesis analysis that takes into account all of the issues associated with the formation and evolution of these systems. We are currently undertaking this type of analysis (Nelson and Rappaport 2003b).

This research was supported in part by NASA under ATP grants NAG5-8368 and NAG5-12522 (to SAR). One of us (LAN) is supported by the Canada Research Chairs program and would like to acknowledge the financial support of the Natural Sciences and Engineering Research Council (NSERC) of Canada. We would also like to thank Jonathan Benjamin and Ernest Dubeau for their technical assistance, and Philipp Podsiadlowski for very helpful discussions.

*Note Added in Proof.–* Markwardt, Smith & Swank (2003) have recently reported the discovery of the fourth-known accreting millisecond pulsar designated as XTE J1807-294. This source was observed during RXTE PCA monitoring of the Galactic-center region. A very significant sinusoidal modulation of the pulse frequency was observed with a period of $35\pm3$ minutes. If this is the orbital period, then XTE J1807-294 would be the shortest-known orbital period of the ultracompact binary millisecond pulsars, and it is likely to have properties very similar to those of XTE J0929-314 and XTE J1751-305.